\documentclass[12pt,preprint,flushrt]{aastex}
\usepackage{rotating,graphicx}
\usepackage{amsmath}
\usepackage{color}
\usepackage{natbib}
\usepackage{soul}
\newcommand{\p}{\partial}

\newcommand{\half}{{\textstyle{1\over2}}}

\def\vnabla{\mbox{\boldmath $\nabla$}}
\def\ffrac#1#2{{\textstyle\frac{#1}{#2}}}

\newcommand{\vx}{{\bf x}}
\newcommand{\vy}{{\bf y}}

\newcommand{\vn}{{\bf n}}

\newcommand{\vE}{{\bf E}}

\newcommand{\ve}{{\bf e}}
\newcommand{\vrr}{{\bf r}}

\newcommand{\msun}{\,M_\odot}
\newcommand{\yr}{\,\mbox{yr}}

\begin{document}

\title{The statistical mechanics of self-gravitating Keplerian disks}

\author{Jihad Touma\altaffilmark{1,2} and Scott Tremaine\altaffilmark{2}}
\altaffiltext{1}{Department of Physics, American University of Beirut,
PO Box 11--0236, Riad El-Solh, Beirut 1107 2020, Lebanon; 
jihad.touma@aub.edu.lb} 
\altaffiltext{2}{School of Natural Sciences, Institute for Advanced
Study, Einstein Drive, Princeton, NJ 08540, USA; tremaine@ias.edu}

\begin{abstract} 

\noindent
  We describe the dynamics and thermodynamics of collisionless particle disks orbiting a massive central body, in the case where the disk mass is small compared to the central mass, the self-gravity of the disk dominates the non-Keplerian force, and the spread in semi-major axes is small. We show that with plausible approximations such disks have logarithmic two-body interactions and a compact phase space, and therefore exhibit thermodynamics that are simpler than most other gravitating systems, which require a confining box and artificial softening of the potential at small scales to be thermodynamically well-behaved. We solve for the microcanonical axisymmetric thermal equilibria and demonstrate the existence of a symmetry-breaking bifurcation into lopsided equilibria. We discuss the relation between thermal and dynamical instability in these systems and draw connections to astrophysical settings, as well as to the wider subject of the statistical mechanics of particles with logarithmic long-range interactions, such as point vortices in two-dimensional fluids.

\end{abstract}

\maketitle 

\section{Introduction}

\noindent
The thermodynamics of isolated, bound, self-gravitating stellar systems (systems of $N$ point particles interacting only through gravitational forces) is notoriously pathological. The infinite volume of physical space forbids maximum-entropy states and the short-range singularity in the potential makes for an unbounded energy, hence undermining the construction of microcanonical ensembles (for reviews see \citealt{pad90} and \citealt{katz03}). To make progress, the usual approach is to introduce artificial cutoffs by confining the N-body system in a spherical box, and/or regularizing or ``softening'' the short-range singularity. With these artifices, canonical and microcanonical equilibria of the self-gravitating gas can be constructed \citep{dlb68,th70,dlb77}. There are regimes with negative heat capacity in the microcanonical equilibria, and these are associated with phase transitions of the corresponding canonical equilibrium (\citealt{th70,dlb77,katz78}; see \citealt{chavanis06} for a relatively recent review). Even simpler toy models are constructed with a view to isolating and analyzing what is thought to be generic behavior, e.g., the Hamiltonian Mean Field (HMF) \citep{ar95}, the Self-Gravitating Ring (SGR) model \citep{sota01}, slab models in which point particles are replaced by infinite sheets \citep{ryb71,jm10,sch13}, and cylindrical models in which point particles are replaced by infinite wires \citep{sto63,ost64,aly94,aly99}. A large body of  literature has evolved around these models, studying their equilibria, phase transitions, dynamical stability and metastability, and connections to the actual systems they are meant to model \citep[e.g.,][]{cdr09}. Although instructive and elegant, these models leave one with the nagging question of what all of this has to do with actual self-gravitating systems in the real world. What remains at the end of the day are robust results on the thermodynamics of artificially imprisoned and mutilated self-gravitating systems, more tentative and largely numerical results on the evolution of realistic systems \citep{bt08}, and heuristic rules relating the properties of the former to the latter.

\section{The Keplerian disk and ring}

\noindent
Here, we bridge the gap between tractable and realistic self-gravitating systems by examining a system that arises naturally (in the study of protoplanetary disks, stellar disks around supermassive black holes, etc.), and in which the pathologies of systems with long-range interactions are naturally resolved.  We start with an infinitesimally thin disk composed of $N\gg1$ identical point particles, each of mass $m$. The particles orbit a central point mass $M_\star\gg Nm \equiv M_{\rm disk}$. The disk is flat; nevertheless particles may orbit in the prograde or retrograde direction (i.e., the inclinations are 0 or $180^\circ$)\footnote{It might seem more natural to model a disk containing only prograde orbits. However, the orbit-averaged gravitational torque on an eccentric orbit does not approach zero as the eccentricity approaches unity.  Thus if the phase space is restricted to prograde orbits, there will be a loss of particles through the boundary at zero angular momentum or $e=1$.}. Since $M_{\rm disk}\ll M_\star$, the particle orbits are nearly Keplerian.  In such disks the dominant relaxation process is resonant relaxation \citep{rt96}, involving secular interactions between particles that cause the orbits to evolve on time-scales of order $M_\star/M_{\rm disk}$ times the orbital period. Relaxation can be studied by averaging over the fast orbital time-scale (the orbital period), that is, replacing each particle by a so-called Gaussian wire (a closed wire following the Keplerian orbit, with linear mass density inversely proportional to velocity). In interactions of this kind the angular momenta or eccentricities of the wires relax, but their energies or semi-major axes do not.  Since each wire has a constant semi-major axis, it is completely specified by its mass $m$, sense of rotation $s$ ($+1$ for prograde and $-1$ for retrograde), eccentricity $e$, and azimuth of periapsis $\varpi$, or instead of the last two the eccentricity vector $\ve\equiv (k,h)\equiv e(\cos\varpi,\sin\varpi)$, which points towards periapsis\footnote{For retrograde particles, our definition of $\varpi$ differs from the usual convention (because $\varpi$ is always measured counter-clockwise from the origin of azimuth rather than in the direction of orbital motion); the advantage of our convention is that prograde and retrograde orbits with the same eccentricity vector occupy the same locus in space.}. The eccentricity vector rotates slowly due to the orbit-averaged force field of the other wires, and varies stochastically due to the even slower effect of resonant relaxation. Note that the conservation of Keplerian energy (semi-major axis) leaves us with a compact $(e,\varpi)$ phase space for the wires to relax in, hence removing the need for artificial confinement.

Alternatively, a particle orbit can be specified by the Poincar\'e variables $\vE \equiv (K,H)\equiv \big(1-\sqrt{1-e^2}\,\big)^{1/2}(\cos\varpi,\sin\varpi)$; these are canonical coordinate-momentum pairs when multiplied by $\sqrt{2m} (G M_\star a)^{1/4}$ (see Appendix \ref{sec:hamxx}). Note that $|\vE|$ and $|\ve|$ both range from 0 to 1, with $\vE\to\ve$ as $|\ve|\to1$ and $\vE\to \ve/\surd 2$ as $|\ve|\to 0$. We shall sometimes call $E=|\vE|$ the Poincar\'e eccentricity, and shift between eccentricity and Poincar\'e eccentricity as needed to keep the formulae as simple as possible. In the models described in this paper, which we call Keplerian rings, all particles are further assumed to share a common (and conserved) semi-major axis $a$. Such a limit is reasonable in disks where the fractional spread in semi-major axes is smaller than the typical orbital eccentricity, but is chosen here mostly for simplicity, as the methods we describe are applicable to disks with any distribution of semi-major axes.

Last but not least, we require the orbit-averaged gravitational potential energy between two particles in the disk: $\Phi(\ve,\ve')=-Gm^2\langle|\vrr-\vrr'|^{-1}\rangle \equiv(Gm^2/a) \phi(\ve, \ve')$, where $\langle\cdot\rangle$ denotes a time average over both orbits (see Appendix \ref{sec:pot}). When eccentricities are small, the averaged potential can be evaluated analytically \citep{bgt83}, $\phi(\ve,\ve')=\phi_L(\ve,\ve')\equiv -4\log 2/\pi+(2\pi)^{-1}\log(\ve-\ve')^2$ plus terms that are $\mbox{O}(e^2,e^2\log e)$. For eccentricities that are not small, $\phi(\ve,\ve')$ must be evaluated numerically by a double integral over the two orbital phases. Most of the calculations described below have been carried out both with the logarithmic potential $\phi_L(\ve,\ve')$ and with a numerical evaluation of $\phi(\ve,\ve')$ on a grid, and the main conclusions are qualitatively and quantitatively unaffected. Therefore for simplicity we present mostly the results with the logarithmic potential\footnote{An alternative approximation is that the potential is logarithmic in the distance between the Poincar\'e eccentricities, $\phi_P=(2\pi)^{-1}\log(\vE-\vE')^2+\mbox{const}$. We have experimented with this approximation and find that the rich behavior described here---bifurcation points, lopsided equilibria, etc.---is present with the exact potential $\phi$ and the approximate potential $\phi_L$ but {\em not} with $\phi_P$.}, except for a brief discussion associated with Figure \ref{fig:four}. 

In the continuum limit, let $n_\pm(\ve)d\ve=f_\pm(\vE)d\vE$ be the number of prograde or retrograde particles on orbits in the eccentricity range $(\ve,\ve+d\ve)$ or $(\vE,\vE+d\vE)$. The total number of prograde and retrograde particles is $n(\ve)\equiv n_+(\ve)+n_-(\ve)$ or $f(\vE)=f_+(\vE)+f_-(\vE)$. In transforming between these we use the relation between phase-space area elements, $d\vE=dKdH=\half dk\,dh/\sqrt{1-e^2}=\half d\ve/\sqrt{1-e^2}$, to write $n_\pm(\ve)=\half f_\pm(\vE)/\sqrt{1-e^2}$. We define a dimensionless mean-field potential of the disk by 
\begin{equation}
\Gamma(\ve)=\frac{1}{N}\int n(\ve')\phi(\ve,\ve') d\ve'=\frac{1}{N}\int f(\vE')\phi(\ve,\ve')d\vE', 
\label{eq:wwqq}
\end{equation}
with  $N =\int n(\ve)\,d\ve=\int f(\vE)\,d\vE$. This potential is the mean-field Hamiltonian, in the sense that (see eq.\ \ref{eq:hameq})
\begin{equation}
\frac{dK}{d\tau}=s\frac{\p \Gamma}{\p H}, \ \frac{dH}{d\tau}=-s\frac{\p \Gamma}{\p K} \ \ \mbox{with}\ \ \tau=\frac{M_{\rm disk}}{2M_\star}\left(\frac{GM_\star}{a^3}\right)^{1/2}\!\!t.
\label{eq:ham}
\end{equation}

The disk's entropy is defined by 
\begin{equation}
S = -\int [f_+(\vE)\log f_+(\vE)+ f_-(\vE)\log f_-(\vE)]\,d\vE. 
\label{eq:entropy}
\end{equation}
Our aim is to extremize the entropy subject to the conservation of the number of particles $N \equiv \int n(\ve)\,d\ve=\int f(\vE)\,d\vE$; the energy $U=\-\half (Gm^2/a)\int n(\ve)n(\ve')\phi(\ve,\ve')\,d\ve\,d\ve'$; and the angular momentum $ L=m\sqrt{GM_\star a}\int [n_+(\ve)-n_-(\ve)]\sqrt{1-e^2}\,d\ve =m\sqrt{GM_\star a}\int [f_+(\vE)-f_-(\vE)](1-E^2)\,d\vE$. We denote the resulting distribution functions and potential $f_\pm^0(\vE)$ and $\Gamma^0(\vE)$.  Using Lagrange multipliers this optimization problem can be solved to give
\begin{align}
f_s^0(\vE)&=\frac{N\alpha}{\beta}\exp[-\beta\Gamma^0(\ve)+s\gamma(1-E^2)], \nonumber \\
n_s^0(\ve)&=\frac{N\alpha}{2\beta\sqrt{1-e^2}}\exp\big[-\beta\Gamma^0(\ve)+s\gamma\sqrt{1-e^2}]\big],
\label{eq:df}
\end{align}
where $\alpha$, $\beta$, $\gamma$ are dimensionless constants and as usual $s=\pm1$ for prograde or retrograde particles. We must have $\alpha/\beta>0$ (the distribution function cannot be negative). The parameter $\beta$ is an inverse temperature, which can be either positive or negative since the phase space is compact. Setting $\Psi(\ve)\equiv \beta\Gamma^0(\ve)$, Poisson's equation (\ref{eq:wwqq}) may now be written
\begin{equation}
\Psi(\ve)=2\alpha\int d\vE'\phi(\ve,\ve')\exp[-\Psi(\ve')]\cosh\gamma\big(1-{E'}^2\big),  
\label{eq:thermal}
\end{equation}
with $\vE=\big(1-\sqrt{1-e^2}\,\big)^{1/2}\ve/e$. This is a nonlinear integral equation for the dimensionless potential $\Psi(\ve)$, whose solution depends on the parameters $\alpha$ and $\gamma$. The inverse temperature $\beta$ is determined from the solution of (\ref{eq:thermal}) by substituting equation (\ref{eq:df}) into the relation $N=\int d\vE\,[f_+(\vE)+f_-(\vE)]$:
\begin{equation}
\beta=2\alpha \int d\vE \exp[-\Psi(\ve)]\cosh\gamma(1-E^2).
\label{eq:nbeta}
\end{equation}

Throughout this paper we shall approximate the potential $\phi(\ve,\ve')$ by the logarithmic potential $\phi_L(\ve,\ve')$, and since $\nabla_\ve^2\phi_L=2\delta(\ve-\ve')$ the integral equation can be replaced by a differential one\footnote{Apart from the factor $\sqrt{1-e^2}$, when $\gamma=0$ this is the equation for the self-gravitating isothermal cylinder \citep{sto63,ost64,klb78,aly94}.},
\begin{equation}
\nabla_\ve^2\Psi=\frac{2\alpha}{\sqrt{1-e^2}}\exp[-\Psi(\ve)]\cosh\gamma\sqrt{1-e^2}.
\label{eq:thermalde}
\end{equation}

Rather than total angular momentum or energy, we shall work with the dimensionless quantities
\begin{align}
\ell&\equiv\frac{L}{Nm\sqrt{GM_\star a}} =\frac{\int  d\vE\,(1-E^2)\exp[-\Psi(\ve)]\sinh\gamma(1-E^2)}{\int d\vE   \exp[-\Psi(\ve)]\cosh\gamma(1-E^2)}, \nonumber \\
u&\equiv\frac{aU}{G(Nm)^2}= \frac{\int d\vE\,d\vE'\,W(\ve)W(\ve')\phi(\ve,\ve')}{2\left[\int d\vE
   \exp[-\Psi(\ve)]\cosh\gamma(1-E^2)\right]^2}, 
\end{align}
where $W(\ve)\equiv\exp[-\Psi(\ve)]\cosh\gamma(1-E^2)$. 
These, together with the integral equation (\ref{eq:thermal}), determine  the potential $\Psi(\ve)$ and the parameters $\alpha$ and $\gamma$, given the conserved quantities $u$ and $\ell$; thus, all thermodynamic equilibria can be parametrized by their dimensionless energy and angular momentum. The dimensionless energy cannot exceed $u=-2\log 2/\pi=-0.44127$, corresponding to particles uniformly distributed on the circle $|\ve|=1$ (see Appendix \ref{sec:axi}); the absolute value of the dimensionless angular momentum cannot exceed unity, and without loss of generality we can restrict $\ell$ to the range $[0,1]$. 

States that are entropy extrema according to equation (\ref{eq:thermalde}) can be either axisymmetric (i.e., depending on $\vE$ only through $E=|\vE|$) or non-axisymmetric. If they are non-axisymmetric the figure is stationary in a frame rotating at the pattern speed given by equation (\ref{eq:pattern}). 

\section{Thermodynamics of the Keplerian ring}

\noindent
We first study axisymmetric entropy extrema, which we construct by solving the differential equation (\ref{eq:thermalde}) (see Appendix \ref{sec:axi}). We plot the results in Figure \ref{fig:one}. The four colored curves show solutions with dimensionless angular momentum $\ell=0,0.5,0.8,0.95$, as functions of the dimensionless energy $u$. The four panels show the mean eccentricity $\langle e \rangle$, fraction of prograde particles, inverse temperature $\beta$, and entropy $S$ (the last of these is for the normalization $N=1$; more generally $S(N)=NS(1)-N\log N$).  Each constant angular-momentum sequence terminates at a point marked by a cross.  Sequences of models with non-zero angular momentum terminate at a mean eccentricity less than unity (as they must, since orbits with $e=1$ have zero angular momentum). The small open triangles in the left part of each panel show the predictions of the low-eccentricity analytic limit (eqs.\ \ref{eq:ost}--\ref{eq:sss}), which agree well with the numerical solutions. Remarkably, the curves of mean eccentricity versus energy (top left panel) almost coincide for the whole range of angular momenta shown.

The axisymmetric entropy extrema become increasingly prograde with increasing energy and mean eccentricity, as they should to maintain a constant angular momentum. The family of axisymmetric solutions includes regions of negative heat capacity ($d\beta/du >0$) and negative temperature $\beta<0$.  The sequence of models with zero angular momentum has $S\to-\infty$ as $\langle e\rangle\to 1$; in this limit the distribution function approaches a singular form in which all the particles have $e=1$. For a given non-zero angular momentum the sequence terminates at finite entropy.

We now investigate the response of these equilibria to small non-axisymmetric perturbations, $\Psi=\Psi^0(e)+\epsilon\psi_m(e)\exp(im\varpi)$, $m>0$, where $\Psi^0(e)$ defines the potential of the unperturbed axisymmetric system (see Appendices \ref{sec:bif} and \ref{sec:saddle}). We substitute this form into the differential equation (\ref{eq:thermalde}) and linearize in the small parameter $\epsilon$. The existence of a solution to the linearized equation implies a bifurcation to a sequence of non-axisymmetric disks that initially have $m$-fold symmetry. We find numerically that (i) bifurcations exist for $m=1$ only; (ii) there is one and only one bifurcation point along the sequence of axisymmetric equilibria at fixed angular momentum $\ell$ for $0\le\ell<0.83356$ (see derivation at the end of Appendix \ref{sec:saddle}), and none for $\ell>0.83356$; (iii) these bifurcations are associated with a transition from entropy maxima, hence thermally stable equilibria, to entropy saddle points which are thermally unstable.  In Figure \ref{fig:one}, we distinguish the regions in which each sequence is stable or unstable by solid and dotted lines, respectively, and mark the locus of bifurcation points by a heavy solid line\footnote{The properties of the system at the bifurcation are continuous functions of the energy in a microcanonical setting. Our preliminary exploration of the thermodynamics of our model disks in the canonical ensemble reveals a richer behavior, including a first-order phase transition at zero angular momentum, which transitions into a second-order transition with increasing angular momentum. The study of the canonical ensemble will take us too far afield in an already lengthy exploration of the microcanonical states and is relegated to future work.}.  The axisymmetric systems are thermally unstable at small mean eccentricity and stable at large mean eccentricity. This result is surprising, since in the limit of small mean eccentricity the equilibrium disks are identical to the isothermal cylinder (eq.\ \ref{eq:thermff}), which is known to be an entropy maximum, and therefore stable (\citealt{klb78,aly94}; see \citealt{aly99} for a generalization to unbounded systems with angular momentum constraint). The explanation is that any isolated system such as the isothermal cylinder is neutrally stable to displacements---in other words, the differential equation governing the density distribution is autonomous---whereas the differential equation (\ref{eq:thermalde}) governing the eccentricity distribution contains terms involving $\sqrt{1-e^2}$ that can make the neutral mode slightly unstable, no matter how small the mean eccentricity.

\begin{figure}[!ht]
\centering
\vspace{-0.4in}
\includegraphics[scale=0.7]{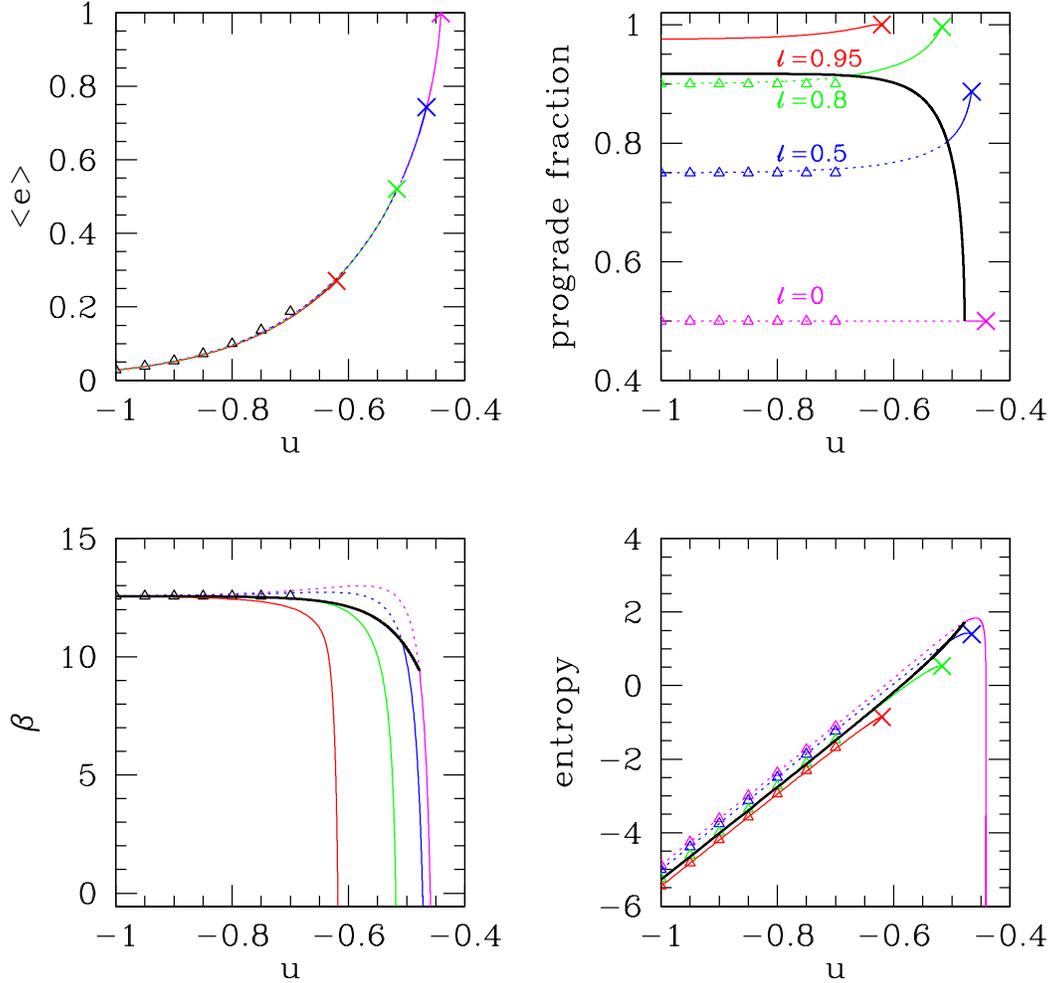}
\vspace{-1.4in}
\caption{\small The properties of axisymmetric disks with a logarithmic potential. Each panel shows systems  with dimensionless angular momentum $\ell=0$ (magenta), 0.5 (blue),  0.8 (green), and 0.95 (red). The four panels show the mean eccentricity (top left), fraction of prograde particles (top right),  inverse temperature (bottom left), and entropy when $N=1$ (bottom  right). Also shown as open triangles are the analytic predictions  for low-eccentricity disks. Each sequence of models terminates at the point marked by a cross. Unstable parts of the constant angular-momentum sequences are represented by dotted lines, while stable parts are shown by solid lines. The locus of bifurcation points, which separates the stable and unstable regions, is shown by a heavy solid line (except in the top left panel, to avoid obscuring the equilibrium sequences with which it almost coincides).  Sequences with $\ell>0.83356$ have no bifurcation.}
\label{fig:one}
\end{figure}

We next construct non-axisymmetric disks using the nonlinear optimization methods described in Appendix \ref{sec:num}. The results are shown in Figure \ref{fig:two}.  The top left and bottom right panels show the same quantities as in Figure \ref{fig:one}. The top right panel shows a measure of the strength of the non-axisymmetry, 
\begin{equation}
I_{\rm max}-I_{\rm min}\equiv \ffrac{5}{2}\sqrt{(\langle k^2\rangle-\langle h^2\rangle)^2+4\langle  kh\rangle^2},
\label{eq:nonaxi}
\end{equation}
where $\langle k^2\rangle=\int d\vE f(\vE) k^2/\int d\vE f(\vE)$, etc.\footnote{It is straightforward to show that $I_{\rm max}-I_{\rm min}$ is the difference between the larger and smaller of the two principal moments of inertia of the disk when the disk has unit semi-major axis and unit mass.  Thus $I_{\rm max}-I_{\rm min}=0$ for axisymmetric disks and $I_{\rm max}-I_{\rm min}=\frac{5}{2}$ for a disk in which all the eccentricity vectors have $e=1$ and are aligned or anti-aligned.} The bottom left panel shows the pattern speed (\ref{eq:pattern}) for the non-axisymmetric disks, defined to be those with $I_{\rm max}-I_{\rm min}>0.01$. 
\begin{figure}[!ht]
\centering
\vspace{-0.7in}
\includegraphics[scale=0.7]{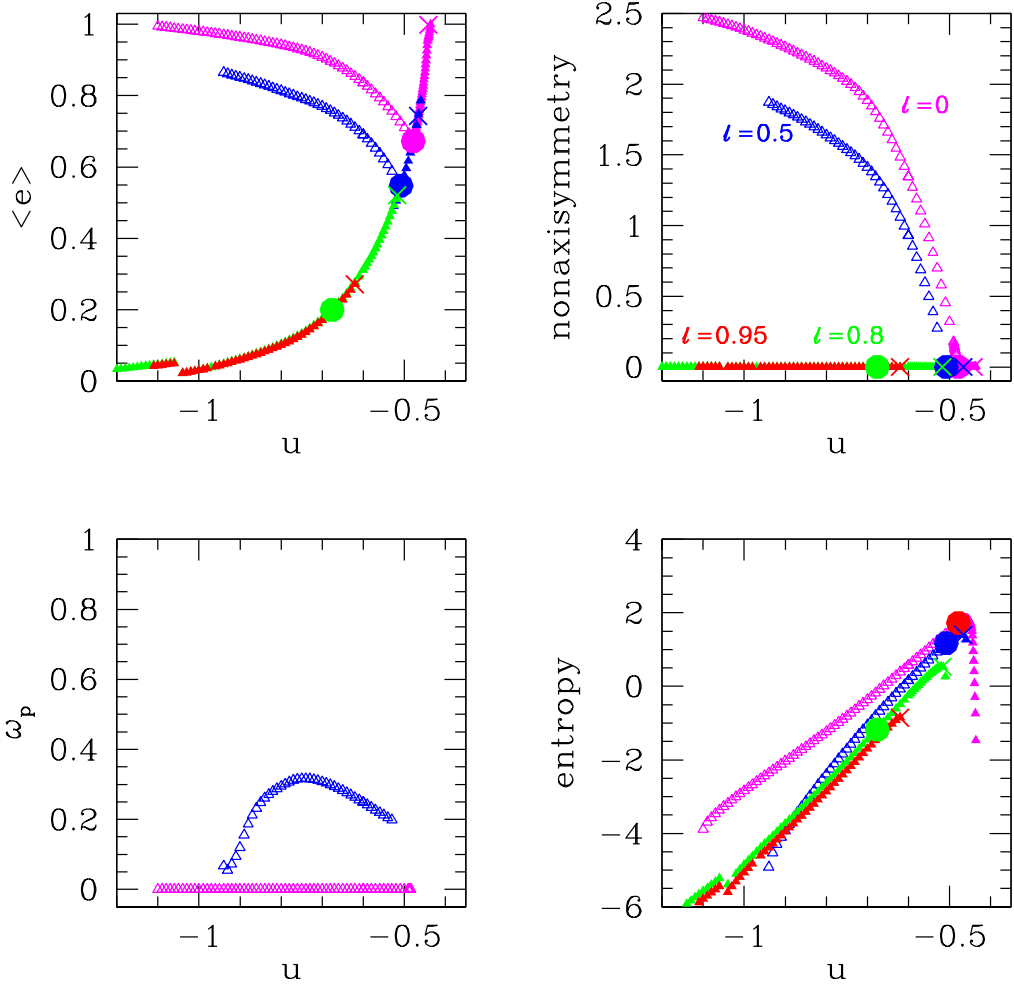}
\vspace{-1.5in}
\caption{\small The properties of axisymmetric and non-axisymmetric disks with a logarithmic potential. Each panel shows systems with dimensionless angular momentum $\ell=0$ (magenta), 0.5 (blue),   0.8 (green), and 0.95 (red). The panels show mean eccentricity (top left), difference $I_{\rm max}-I_{\rm min}$ (eq.\ \ref{eq:nonaxi}) between the major and minor axes of the inertia ellipse (top right), pattern speed (bottom left) and entropy when $N=1$ (bottom right).   The triangles show the maximum-entropy states computed via nonlinear optimization; closed triangles are  axisymmetric ($I_{\rm max}-I_{\rm min}<0.01$) and open triangles are  non-axisymmetric. Crosses denote the termination of the axisymmetric sequences and solid circles denote bifurcation points, both taken from Fig.\ \ref{fig:one}.}
\label{fig:two}
\end{figure}
First consider the magenta triangles, which outline the locus of models with zero angular momentum and equal fractions of prograde and retrograde particles. The models begin near mean eccentricity $\langle e\rangle=1$ and energy $u=-0.44127$, corresponding to an axisymmetric disk composed of radial orbits. As the energy is reduced the models initially follow the axisymmetric sequence shown in Figure \ref{fig:one}. At the bifurcation point, marked by a solid circle at $u=u_b=-0.478$, $\langle e\rangle=0.670$, the models leave the axisymmetric sequence, which is no longer an entropy maximum beyond this point, to follow a non-axisymmetric sequence with growing mean eccentricity. This sequence terminates in a disk composed of particles on radial orbits with aligned eccentricity  vectors, $\langle e\rangle=1$ and $I_{\rm max}-I_{\rm min}=\frac{5}{2}$. The numerical models terminate at $u=-1.1$ due to the limited resolution of our grid but we believe that the non-axisymmetric sequence should extend to $u\to-\infty$.

The behavior of models with angular momentum $\ell=0.5$ (blue triangles) is qualitatively similar, in that the axisymmetric sequence bifurcates to a non-axisymmetric sequence as the energy is decreased (at energy $u_b=-0.508$ and mean eccentricity $\langle e\rangle=0.549$). However, neither the axisymmetric nor the non-axisymmetric sequence can achieve $\langle e\rangle=1$ since such a system would be composed entirely of radial orbits, which have zero angular momentum.  Instead, as the energy $u$ becomes more negative, the orbits cluster more and more tightly around a single value of the eccentricity vector, with magnitude given by $\langle e\rangle=e_f\equiv\sqrt{1-\ell^2}$. The numerical models terminate at $u\simeq-0.94$ but we believe this is because of the limited resolution of our grid, and the sequence should asymptote to a horizontal line at $e_f$ that extends to $u\to-\infty$. 

\begin{figure}[!ht]
\centering
\vspace{-0.7in}
\includegraphics[scale=0.67]{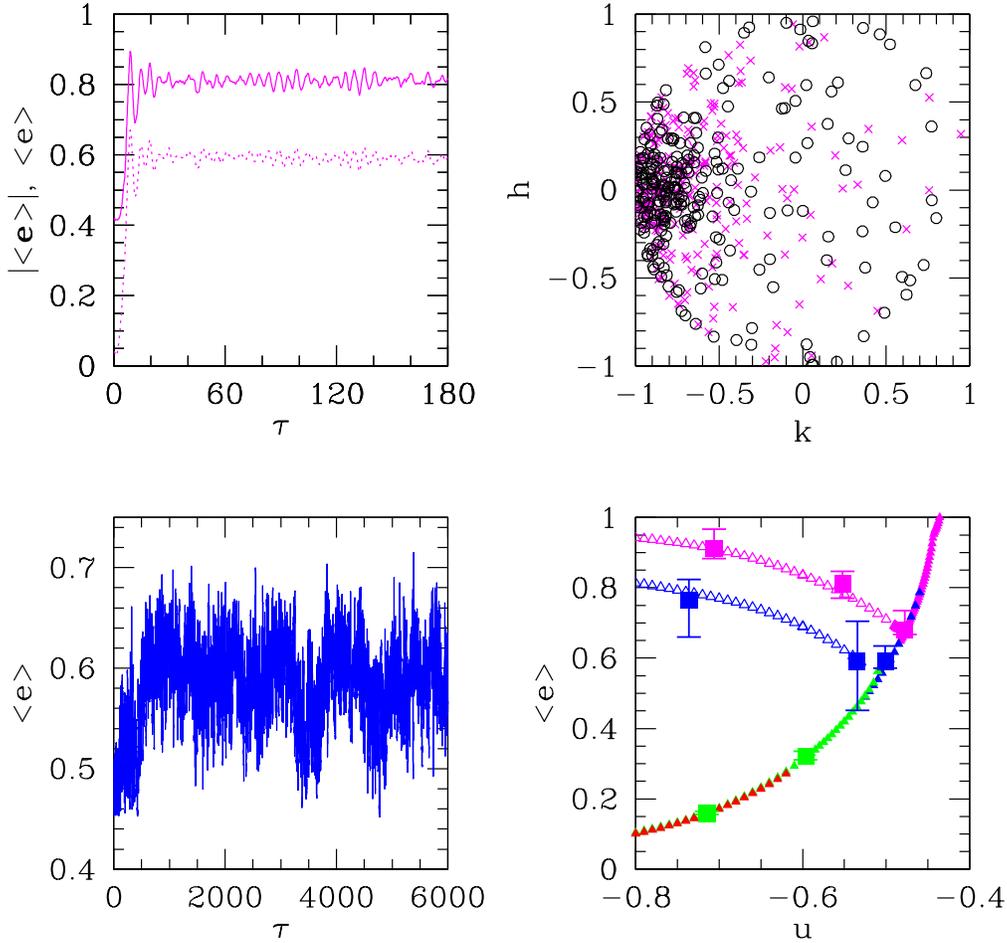}
\vspace{-1.4in}
\caption{\small Top left: the evolution of the mean eccentricity (solid) and the norm of the mean eccentricity vector (dashed) in a 256-wire dynamical simulation of the unstable axisymmetric equilibrium at $\ell=0$ and $u=-0.55$.  Top right: a sample of the non-axisymmetric maximum-entropy equilibrium with this energy and angular momentum (magenta crosses) superposed with the dynamical simulation (black circles) around $\tau=182$. The directions of the mean eccentricity vectors of the two states were reoriented to coincide. Bottom left: the evolution of the mean eccentricity of a 128-wire sample of an unstable axisymmetric equilibrium with $\ell =0.5$ and $u = -0.535$ (the bifurcation energy at $\ell=0.5$ is $u = -0.508$). Bottom right: the final mean eccentricity of simulated ensembles (with error bars), superimposed on a zoom-in of the top left panel of Fig.\ \ref{fig:two}. All initial states that are expected to be dynamically stable stayed close to their initial axisymmetric state for the duration of the simulation. All initial states that were expected to be unstable (except for the one at $\ell=0.8$, $u=-0.715$) became lopsided with a final mean eccentricity close to that of the maximum-entropy states in Fig.\ \ref{fig:two}. 
}
\label{fig:three}
\end{figure}

For $\ell=0.8$ (green triangles), our earlier analysis of the thermal stability of axisymmetric systems implies that there is a bifurcation to a non-axisymmetric sequence at $u_b=-0.675$, $\langle e\rangle=0.200$, but the numerical models remain on the axisymmetric sequence for all energies. This is presumably an artifact of our limited resolution, since (i) a bifurcation point exists only for $\ell<0.83356$, which is close to $\ell=0.8$; (ii) the entropy curves in the bottom right panel of Figure \ref{fig:three} are very close together once $\ell\gtrsim 0.5$ so it is difficult for the optimization code to settle onto the non-axisymmetric sequence. For $\ell=0.95$ no bifurcation is expected or observed.

\section{Thermal and dynamical stability}

\noindent
We come now to the relation between thermal and dynamical instability.  In the orbit-averaged dynamics described here, dynamical instability typically proceeds on the secular time-scale, which is longer than the orbital period $2\pi(a^3/GM_\star)^{1/2}$ by a factor $\sim M_\star/NM=M_\star/M_{\rm disk}$ (i.e., $\tau\sim 1$ in the notation of eq.\ \ref{eq:ham}).  We do not consider possible dynamical instabilities on the time-scale of the orbital period; on this time-scale the disks should be stable since their mass is much smaller than the central mass. Thermal instability proceeds on the resonant relaxation time-scale, which is expected to be longer than the secular time by a factor $\sim N$ \citep{rt96}. Thermal stability implies dynamical stability, but thermal instability need not imply dynamical instability \citep{bart71, ih79} since the collisionless Boltzmann equation conserves phase-space density and the thermal instability may not. For similar reasons, a dynamically unstable initial state does not normally evolve towards a maximum-entropy final state on the secular time. Thus, we expect that dynamical instability leads in a timescale $\tau=\mbox{O}(1)$ to an ``intermediate'' state that is a time-independent solution of the collisionless Boltzmann equation, and that the intermediate state then evolves on a timescale $\tau=\mbox{O}(N)$ to the maximum-entropy state. 

Analyses of the dynamical stability of collisionless near-Keplerian stellar disks, with or without a range of semi-major axes \citep{tou02, ss10, kt13}, generally find that if there is a sufficient number of counter-rotating particles (sufficiently small total angular momentum) the disks are dynamically unstable and settle into lopsided states on a secular time-scale. To determine whether these conclusions apply to the disks studied in this paper, we have solved the linearized collisionless Boltzmann equation for the axisymmetric models shown in Figure \ref{fig:one} (see Appendix \ref{sec:dyn}). We find that dynamical instabilities are present in some models, and the onset of dynamical instability occurs at the same bifurcation points at which the disk becomes thermally unstable and the sequence of non-axisymmetric maximum-entropy models begins (to within 0.3\% in energy $u$). In other words, it appears that the axisymmetric models are dynamically unstable if and only if they are thermally unstable. 

To explore further the relation between dynamical and thermal instability in these systems, we simulated the dynamical evolution of ensembles of Gaussian wires selected from the distribution functions of axisymmetric thermal equilibria. We call these N-wire simulations in analogy to N-body simulations \citep{ttk09}. Ensembles with 128, 256 and 512 wires were simulated at $\ell=0, 0.5$ and $0.8$, at energies above and below the bifurcation points identified in Figure \ref{fig:one}.  In the top left panel of Figure \ref{fig:three}, we display the mean eccentricity and norm of the mean eccentricity vector for an initially axisymmetric system with $\ell=0$ and $u =-0.55$, which is thermally unstable according to Figures \ref{fig:one} and \ref{fig:two}. The mean eccentricity shows a rapid departure from its equilibrium value in the axisymmetric system ($\langle e\rangle=0.42$) through an (overstable) cycle, which saturates after a sequence of oscillations of gradually decreasing amplitude at a mean eccentricity $\langle e\rangle= 0.81$. The mean eccentricity vector follows suite, departing from $\langle\ve\rangle=0$ and saturating in a lopsided configuration with $|\langle\ve\rangle|\simeq 0.59$. We identify these configurations with the ``intermediate'' equilibria described above. On timescales $\tau=\mbox{O}(N)$ we expect that the intermediate equilibria should evolve toward maximum-entropy equilibria. We have not been able to detect this evolution, simply because the macroscopic properties of the intermediate equilibria are already close  to those of the maximum-entropy equilibria when they first appear. For example, the mean eccentricity and mean eccentricity vectors in the intermediate state at $\tau=20$--40 ($\langle e\rangle=0.81$ and $|\langle \ve\rangle|= 0.59$) 
are within a few percent of the corresponding quantities in the maximum-entropy state with the same energy and angular momentum ($\langle e\rangle=0.78$ and $|\langle \ve\rangle|= 0.61$).
Similarly, the non-axisymmetry parameter $I_{\rm max}-I_{\rm min}$ (eq.\ \ref{eq:nonaxi}) fluctuates around 1.0 in the simulation, close to but 10\% larger than its value of 0.91 in the maximum-entropy state.  In the top right panel of Figure \ref{fig:three}, we superpose the eccentricity vectors of the 256 wires in the N-wire simulation (circles) at $\tau=182$ onto a 256-point sample of the eccentricity vectors in the maximum-entropy state with the same energy and angular momentum (magenta crosses). The distributions are similar, but the mean eccentricity vector of the maximum-entropy state is smaller and its spread around the mean is broader.

We have carried out N-wire simulations of zero angular momentum ($\ell=0$) axisymmetric equilibria at other energies, both below and above the bifurcation value  $u_b = -0.478$, and these were equally robust in converging in the mean to states close to the expected maximum-entropy states of Figure \ref{fig:two}. The case $\ell=0.5$ is more complex. As expected, the N-wire simulations showed stability and instability for values of the energy larger and smaller, respectively, than the bifurcation energy $u_b = -0.508$. However the dynamical evolution was far more tortuous.  In the bottom left panel of Figure \ref{fig:three}, we follow the mean eccentricity of an ensemble of 128 wires sampling an initially axisymmetric (and thermally unstable) equilibrium with $\ell=0.5$ and $u= -0.533$. The cluster transitions rather fast to a lopsided state with mean eccentricity $\langle e\rangle\simeq 0.53$ then undergoes a further transition around $\tau \sim 500$ to a more lopsided state with $\langle e\rangle\simeq 0.6$, almost exactly the value in the maximum-entropy state ($\langle e\rangle=0.59$). In both states the mean eccentricity exhibits fluctuations with an amplitude of about 0.08. By $\tau =5000$, the lopsided system is precessing with a mean pattern speed  $\omega_p = 0.22$, close to the value $\omega_p = 0.20$ expected in the maximum-entropy state with the same energy and angular momentum.  The evolution over nearly 200 mode precession periods ($\tau=6000$) shows a number of intermittent transitions to states with lower mean eccentricity and few signs of settling down to a maximum-entropy configuration; in general states with lower mean eccentricity have higher pattern speeds and vice versa. A larger N-wire simulation ($N=256$) showed similar transitions over the same time-scale so these are unlikely to be an artifact of small $N$. A simulation with $\ell=0.5$ and $u=-0.735$, further from the bifurcation energy, lingers around the nearly axisymmetric initial state until about $\tau=2000$, before it undergoes a series of transitions to larger mean eccentricity,  eventually (by $\tau=7500$) attaining  $\langle e \rangle \simeq 0.77$ (with fluctuations of about 0.15), close to the mean eccentricity of the maximum-entropy state ($\langle e \rangle =0.78$). The pattern speed settled after a series of ups and downs to a mean value $\omega_p = 0.29$, close to the value $\omega_p = 0.31$ expected in the maximum-entropy state with the same energy and angular momentum. Models with $\ell=0.8$ revealed in dynamical simulations some of the same pathologies displayed by their counter-parts in the search for maximum entropy non-axisymmetric states: in particular, states that are predicted to go unstable seemed stuck in the neighborhood of their initial near-equilibrium configuration, even in relatively lengthy simulations with $N=256$ wires. The final states of all of these simulations are displayed as solid squares with error bars in the bottom right panel of Figure \ref{fig:three}), along with the maximum-entropy equilibria shown in the top left panel of Figure \ref{fig:one}. 

We conclude that in some of our models dynamical instability leads to ``intermediate'' states that are close to maximum-entropy states; other models, particuarly those with significant angular momentum, often seem to linger in, or oscillate between, metastable states. Possibly this behavior is associated with the small difference in entropy between the axisymmetric and non-axisymmetric entropy extrema (compare the lower-right panels of Figures \ref{fig:one} and \ref{fig:two}). 

\section{Discussion}

\noindent
We have examined the maximum-entropy states of a razor-thin disk of collisionless masses orbiting a massive central body. The disks may contain particles on both prograde and retrograde orbits and particles are allowed to flip between prograde and retrograde orbits, but the total energy and angular momentum of the disk are conserved. The disk mass is assumed to be much smaller than the mass of the central body, so the interaction potential between two particles can be approximated by its orbit-averaged value. This approximation is appropriate if the disk age is shorter than the time-scale for two-body relaxation due to close encounters. The orbit-averaged interaction between particles leads to resonant relaxation, in which the angular momenta  and eccentricities of the particles relax, but the semi-major axes remain fixed. For simplicity, we focus in this paper on the somewhat artificial case in which all the particles have the same semi-major axis (``Keplerian rings''), although our methods are easily adapted to more general disk models. 

Although the Keplerian rings described here are artificial systems intended mainly as aids in exploring the dynamics and statistical mechanics of self-gravitating stellar systems, it is useful to relate them to the properties of a real astrophysical system to which they may offer insight. The center of the Milky Way galaxy contains a black hole surrounded by a near-Keplerian stellar system, with the following properties (taken from \citealt{kt11}): black-hole mass $M_\bullet=4\times 10^6\msun$; number of stars within 0.1 pc $N=5\times 10^4$; orbital period at 0.1 pc $1.5\times10^3\yr$; age $\sim 10^{10}\yr$; resonant relaxation time $\sim 5\times10^7\yr$. 

We construct the maximum-entropy equilibria that should be the end-state of resonant relaxation. The natural expectation is that such disks should be axisymmetric, with an eccentricity distribution given approximately by the analytic solution in \citep{sto63, ost64}, at least so long as the mean eccentricity is not too large. This expectation is not correct: for a given angular momentum we find that the maximum-entropy state has a minimum mean eccentricity (top left panel of Figure \ref{fig:two}) which is achieved at a critical value of the energy, $u_b$. The maximum-entropy state is axisymmetric for energy $u > u_b$ and lopsided for $u<u_b$. For $u<u_b$ the axisymmetric equilibrium is an entropy extremum but not a maximum. Both the pattern speed and the temperature of the lopsided disks are generally positive. Essentially, as the disk is cooled to lower and lower energies the stellar orbits concentrate around a single eccentricity vector $\ve_0$ whose magnitude is determined by the angular momentum, $e_0=\sqrt{1-\ell^2}$. 

\begin{figure}[!ht]
\centering
\vspace{-0.7in}
\includegraphics[scale=0.7]{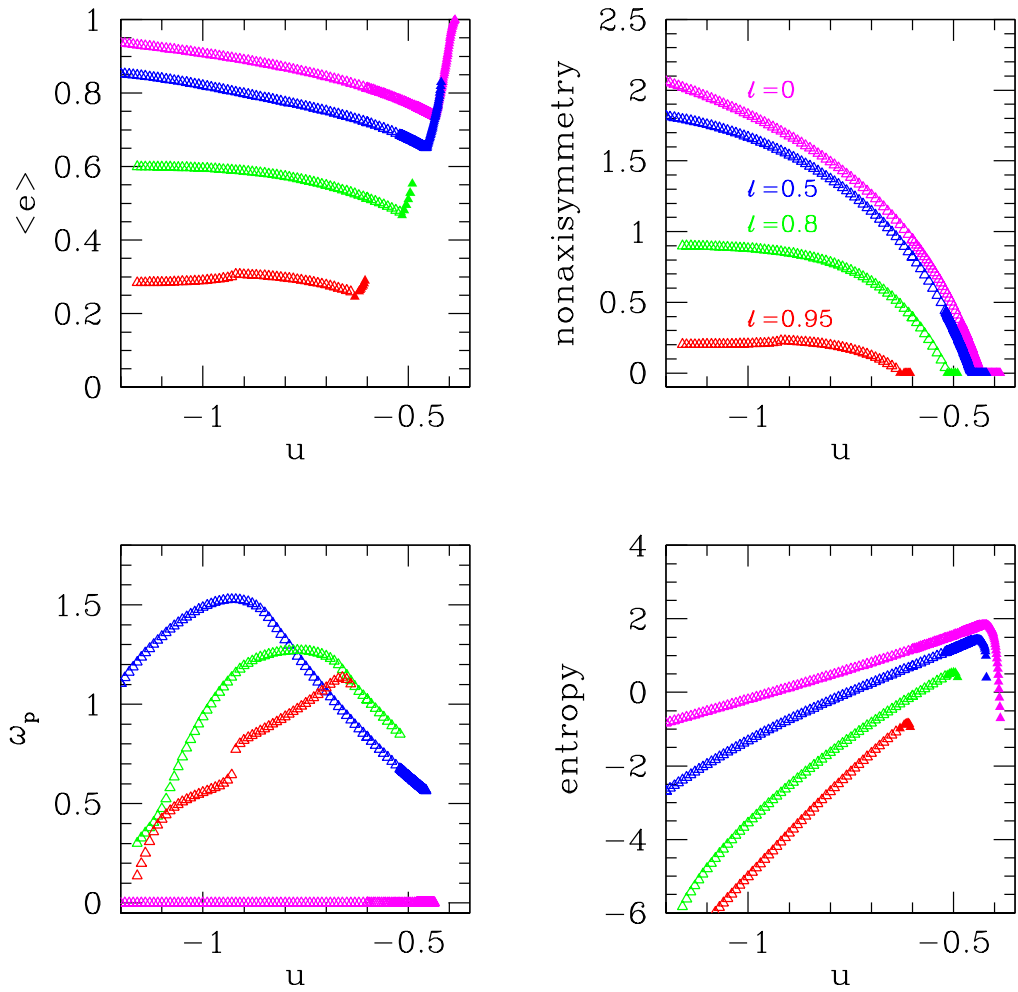}
\vspace{-1.5in}
\caption{\small As in Figure \ref{fig:two}, except the gravitational potential is computed using the exact expression (\ref{eq:pot}) rather than the logarithmic approximation (\ref{eq:smalle}). }
\label{fig:four}
\end{figure}

The results presented in this paper are based on a logarithmic approximation to the orbit-averaged potential energy between two particles, an approximation that is valid only in the limit of small eccentricities. The mean eccentricities of the non-axisymmetric equilibria are large enough to cast doubt on the validity of this approximation. However, we have repeated our calculations using the exact orbit-averaged potential (computed on a three-dimensional grid in $\ve_1$--$\ve_2$) and we found that the maximum-entropy states produced with the logarithmic potential and the exact potential have all the same qualitative features (bifurcation points, minimum mean eccentricity, lopsided equilibria, etc.). Maximum-entropy models computed in this way are shown in Figure \ref{fig:four}, which should be compared with Figure \ref{fig:two}. 
 
The numerical methods we have used need to be improved. At present we find the entropy maxima using sequential quadratic programming, defining the distribution function on a 4096-point grid in eccentricity space. In a few cases we find suspicious numerical artifacts (e.g., the small discontinuity near $u=-1.05$, $\langle e\rangle=0.03$ in the top left panel of Figure \ref{fig:two}), and in most cases convergence is quite slow. This said, we have confirmed our main results with  Markov-chain Monte Carlo simulations, basis-function expansions of the central integral equation (see Appendix \ref{sec:basis}), numerical solutions of the analogous differential equation (\ref{eq:thermalde}) for axisymmetric states, and nonlinear optimization using smaller grids.

We have also examined possible dynamical instabilities in axisymmetric Keplerian rings, which are expected to occur on the secular times-scale, that is, a times-scale longer than the orbital period by the ratio of the central mass to the disk mass. We find that the rings are dynamically unstable if and only if they are thermally unstable.  We showed via N-wire simulations that dynamical instability in these disks produces lopsided states. We observe that in some but not all of our experiments these are close to the maximum-entropy solutions in mean eccentricity, distribution of eccentricity vectors, and precession rates or pattern speeds; we do not have an explanation for this similarity nor do we know whether it has an illuminating physical explanation.  The presence of this instability is in line with earlier findings of generic dynamical instabilities in disks containing a retrograde stellar population \citep{tou02,ttk09,gss12}\footnote{In both N-wire and N-body simulations of unstable counter-rotating disks \citep{ttk09, kt13}, stellar orbits experience large-amplitude oscillations in inclination when their eccentricity increases beyond a critical value. Such eccentricity-inclination instabilities may operate in the disks we consider here if given the freedom to do so,  and make it imperative to generalize our results to three-dimensional maximum-entropy equilibria. In addition to endowing our models with greater physical realism, the extra degree of freedom provides a natural way to resolve the otherwise singular transition from the prograde to the retrograde sector of phase space.}. However, we do not know why the final state of the dynamical instability is so similar to the maximum-entropy state resulting from thermal instability, since this is not generally true in self-gravitating systems (e.g., in collapse of spherical systems that are not initially in virial equilibrium, where there {\em is} no maximum-entropy state)---perhaps part of the answer is that the phase space of the systems examined here is compact. 

Our results on the equilibria of self-gravitating systems with logarithmic two-body potentials in eccentricity space have a strong kinship with the far more extensive body of work on the statistical mechanics of point vortices in compact domains (see Appendix \ref{sec:vortex}). The interaction potential for vortices is logarithmic in physical space, so physical space for vortices maps into eccentricity space for wires, and conserved circulation in point vortices to conserved semi-major axes in the secular dynamics of wires. Of course there are obvious and important differences: in self-gravitating wires the potential energy does not depend on the direction of motion---prograde or retrograde---of the particles, whereas it does depend on the sign of circulation of vortices; wires can evolve between prograde and retrograde, while vortices cannot change their circulation; negative-temperature states in vortices are prone to phase transitions, whereas negative-temperature rings appear perfectly stable in axisymmetric configurations. This said, much of the analytic machinery developed to study the existence and stability of solutions in the point vortex case should extend quite naturally to our problem.

We have found remarkable and unexpected complexity in the thermodynamics of near-Keplerian stellar disks. These results are of interest both for exploring the thermodynamics of systems with long-range forces and because they suggest that many near-Keplerian, nearly collisionless, astrophysical disks (disks near supermassive black holes, debris disks around young stars, etc.) may naturally develop a lopsided configuration. 

\acknowledgments

This research was supported in part by NASA grant NNX11AF29G. JT acknowledges the support
of an Arab Fund Research Fellowship for the year 2013--2014, which allowed him an extended stay at the IAS and a briefer one at the IHP (Paris), the hospitality of both institutes being greatly appreciated. 

\appendix

\section{The Keplerian ring}

\noindent
We assemble here the mathematical machinery, remarks, and results which underlie and amplify the results and assertions in the body of the text.

\subsection{The interaction potential}

\label{sec:pot}

\noindent 
The study of secular dynamics requires the time-averaged gravitational interaction energy between two particles:
\begin{equation}
\Phi(a_1,\ve_1,a_2,\ve_2)=-Gm^2\bigg\langle \frac{1}{|\vrr_1-\vrr_2|}\bigg\rangle
\end{equation}
where $a_{1,2}$ and $\ve_{1,2}$ are the semi-major axes and eccentricity vectors of the particles, and $\langle\cdot\rangle$ denotes a time average over both orbits. In this paper we examine the special case where all particles share the same semi-major axis $a$. Then 
\begin{equation}
\Phi(a,\ve_1,a,\ve_2)=\frac{Gm^2}{a}\phi(\ve_1,\ve_2)
\end{equation}
where 
\begin{align}
\label{eq:pot}
  \phi(\ve_1,\ve_2)&=\phi(e_1,e_2,\varpi_1-\varpi_2) \\
  =-&\frac{1}{(1-e_1^2)^{1/2}(1-e_2^2)^{1/2}}\int_0^{2\pi}\frac{df_1}{2\pi}
  \int_0^{2\pi}\frac{df_2}{2\pi}
  \frac{r_1^2r_2^2}{[r_1^2+r_2^2-2r_1r_2\cos(f_1+\varpi_1-f_2-\varpi_2)]^{1/2}};\nonumber 
\end{align}
here $f_i$ and $\varpi_i$ are the true anomaly and longitude of periapsis of particle $i$ (so $\ve=e\cos\varpi \hat\vx + e\sin\varpi\hat\vy$), and 
\begin{equation}
  r_i=\frac{1-e_i^2}{1+e_i\cos f_i}.
\end{equation}
The function in equation (\ref{eq:pot}) is symmetric in its arguments,
$\phi(\ve_1,\ve_2)=\phi(\ve_2,\ve_1)$, and is rotationally invariant,
that is, it depends on $\varpi_1$ and $\varpi_2$ only through
$\Delta\varpi=\varpi_2-\varpi_1$.  When the
eccentricities are small, the integral can be evaluated analytically,
\begin{equation}
  \phi(\ve_1,\ve_2)=\phi_L(\ve_1,\ve_2) +\mbox{O}(e^2,e^2\log e)\quad\mbox{where}\quad \phi_L(\ve_1,\ve_2)\equiv -\frac{4\log 2}{\pi} +\frac{1}{2\pi}\log(\ve_1-\ve_2)^2.
  \label{eq:smalle}
\end{equation}

When $e_1=0$, the potential can be expanded in powers and logarithms of $e_2$,
\begin{align}
  \phi(0,e_2)=&-\frac{4\log
  2}{\pi} +\bigg(\frac{3\log 2}{8\pi}-\frac{5}{64\pi}\bigg)e_2^2 +
\bigg(\frac{165\log 2}{2048\pi}-\frac{1187}{32768\pi}\bigg)e_2^4
+\bigg(\frac{525\log 2}{16384\pi} \\
&-\frac{21635}{786432\pi}\bigg)e_2^6
+\frac{1}{2\pi}\log
e_2^2\bigg(1-\frac{3}{32}e_2^2-\frac{165}{8192}e_2^4-\frac{525}{65536}e_2^6\bigg)
  +\mbox{O}(e_2^8,e_2^8\log e_2^2). \nonumber 
\end{align}
The integral for $\phi(\ve_1,\ve_2)$ diverges logarithmically
as $\ve_2\to\ve_1$, suggesting that the potential can be written in
the form 
\begin{equation}
\phi(\ve_1,\ve_2)=\phi_a(e_1^2,e_2^2,\ve_1\cdot\ve_2)+\phi_b(e_1^2,e_2^2,\ve_1\cdot\ve_2)\log(\ve_1-\ve_2)^2
\label{eq:potapp}
\end{equation}
where $\phi_a$ and $\phi_b$ are smooth functions. The following
functional forms fit the potential with an rms fractional error of
2\%:
\begin{align}
\phi_a&=-0.91157 + 0.22230 (e_1^2 + e_2^2) -
0.32828e_1e_2\cos\Delta\varpi
+ 0.10986e_1^2e_2^2 \nonumber \\
&\qquad\quad - 0.14496 (e_1^4 + e_2^4) 
 + 0.10428 (e_1^2 + e_2^2) e_1e_2\cos\Delta\varpi + 
 0.098476 e_1^2 e_2^2\cos^2\Delta\varpi, \nonumber \\
 \phi_b&=0.14468 + 0.050327 (e_1^2 + e_2^2) + 0.21318 e_1e_2\cos\Delta\varpi.
\label{eq:fit}
\end{align}
The numerical experiments in this paper use the logarithmic potential $\phi_L$ (eq.\ \ref{eq:smalle}) even for eccentricities of order unity, where it is not strictly valid. The reason for this is that we have also conducted experiments with the accurate averaged potential\footnote{These experiments do not use equation (\ref{eq:fit}) or other fitting formulae; instead they rely on numerical evaluations of the double integral (\ref{eq:pot}) on a uniform $16^3$ or $32^3$ grid in the space $(E_1,E_2,\Delta\varpi)$ where $E_i=(1-\sqrt{1-e_i^2})^{1/2}$ is the Poincar\'e eccentricity.}, and found that these are in essential qualitative agreement with those obtained via the logarithmic potential (see Figure \ref{fig:four}). 

\subsection{Hamilton's equations}

\label{sec:hamxx}

\noindent
The Hamiltonian of a particle with eccentricity $\ve_i$ is 
\begin{equation}
\widetilde \Gamma(\ve_i)\equiv \frac{Gm^2}{a}\sum_{j=1}^N\phi(\ve_i,\ve_j).
\end{equation}

The Poincar\'e variables $\widetilde\vE_i\equiv (\widetilde K_i,\widetilde H_i)\equiv
(2m)^{1/2}(GM_\star a)^{1/4}\big(1-\sqrt{1-e_i^2}\big)^{1/2}(\cos\varpi_i,\sin\varpi_i)$ are a
canonical coordinate-momentum pair, that is, they evolve at a rate 
 \begin{equation} 
   {d \widetilde K_i\over dt}=s_i{\p \widetilde\Gamma\over\p \widetilde H_i},\qquad 
   {d \widetilde H_i\over dt}=-s_i{\p \widetilde\Gamma\over\p \widetilde K_i}.
 \end{equation} 
For prograde particles $\widetilde K$ is the coordinate and $\widetilde H$ is the momentum, while their roles are reversed for retrograde particles. We rescale variables, time and Hamiltonian:
 \begin{align}
   \vE_i\equiv& (K_i,H_i)\equiv
  {\textstyle \big(1-\sqrt{1-e_i^2}\,\big)^{1/2}}(\cos\varpi_i,\sin\varpi_i), \nonumber \\
   \tau=&\frac{M_{\rm disk}}{2M_\star}\left(\frac{GM_\star}{a^3}\right)^{1/2}\!\! t\nonumber \\
   \Gamma(\ve)\equiv &\frac{a\widetilde\Gamma(\ve)}{N Gm^2}=\frac{1}{N}\sum_{i=1}^N\phi(\ve,\ve_i).
   \label{eq:dimen}
 \end{align}
In these variables, Hamilton's equations read 
 \begin{equation} 
   {d  K_i\over d\tau}=s_i{\p \Gamma\over\p  H_i},\qquad 
   {d  H_i\over d\tau}=-s_i{\p \Gamma\over\p  K_i}.
\label{eq:hameq}
 \end{equation}
The analogous equations for $\ve_i=(k_i,h_i)=e_i(\cos\varpi_i,\sin\varpi_i)$ are
\begin{equation} 
   {d k_i\over d\tau}=2s_i{\textstyle\sqrt{1-e_i^2}}\frac{\p \Gamma}{\p h_i},\qquad 
   {d h_i\over d\tau}=-2s_i{\textstyle \sqrt{1-e_i^2}}\frac{\p \Gamma}{\p k_i}.
\label{eq:hkdot}
\end{equation}

There are two conserved quantities: the energy
and the angular momentum
\begin{align}
    U=&\frac{Gm^2}{2a}\sum_{i,j\atop
      i\not=j}\phi(\ve_i,\ve_j)=\frac{GNm^2}{2a}\sum_i\Gamma(\ve_i), 
\nonumber \\  L=&m\sqrt{GM_\star a}\sum_{i=1}^N
s_i\sqrt{1-e_i^2}=m\sqrt{GM_\star a}\sum_{i=1}^N
s_i\big(1-E_i^2\big).
\label{eq:int}
\end{align}

The phase-space area element is
\begin{equation}
         d\vE=dKdH=\frac{dk\,dh}{2\sqrt{1-e^2}}=\frac{d\ve}{2\sqrt{1-e^2}}.
\label{eq:area}
\end{equation}

It is sometimes useful to think of the phase space as the surface of a sphere
of unit radius, in which the azimuthal angle is $\varpi_i$ and
the polar angle is $0\le\theta\le\pi$ where
\begin{equation}
\sin\theta=e, \quad \cos\theta=s\sqrt{1-e^2}.
\label{eq:thetadef}
\end{equation}
We call this the eccentricity sphere. The northern (southern)
hemisphere represents prograde (retrograde) orbits, and the
equator represents radial orbits ($e=1$).  The area element on
the eccentricity sphere is proportional to the area element in phase
space, $d^2\Omega=2 dKdH$. 

\subsection{Relation to point vortices}

\label{sec:vortex}

\noindent Equations (\ref{eq:hkdot}) are closely related to Kirchhoff's equations for the motion of Helmholtz point vortices on the sphere \citep{kies12}. Let the unit vector ${\vn} = (x,y,z)=(k, h, s\sqrt{1-h^2-k^2})$ denote the location of one of our particles on the eccentricity sphere. The rate of change of $x$ and $y$ is given by equations (\ref{eq:hkdot}). Taking the derivative of $z$ with respect to $\tau$ we get 
\begin{equation}
   {d z\over d\tau}=  2 y \frac{\p \Gamma}{\p x} -2 x\frac{\p \Gamma}{\p y}.
\label{eq:ndot}
\end{equation}
The equations of motion can be rewritten in the compact vectorial form
\begin{equation} 
   {d {\vn}\over d\tau}=  -2 {\vn} \times {\vnabla}_{{\vn}} \Gamma.
\label{eq:vndot}
\end{equation}
While we believe that this formulation offers greater insight and elegance, and may well simplify 
the mathematical analysis of $N$-wire systems, all the calculations in this paper are based on the (mathematically equivalent) geometry of a double copy of a disk of unit radius, corresponding to populations of prograde and retrograde wires.

Equation (\ref{eq:vndot}) also governs the dynamics of a collection of point vortices all having the same circulation, evolving according to a Helmholtz-like Hamiltonian that is proportional to $\Gamma$.  The only difference between the Helmholtz Hamiltonian and ours (apart from a constant of proportionality) is that Helmholtz's is proportional to $-\log(\vn_i-\vn_j)^2$ whereas ours is proportional to $\log(\ve_i-\ve_j)^2$ (eq.\ \ref{eq:smalle}). 

More generally, our equations, methods and to some extent solutions have strong kinship with the vast body of literature dedicated to the statistical mechanics of point vortices  \citep{onsager49, jm74, kida75, miller90, rs91},  and of guiding-center plasmas \citep{jm73, so90} in compact planar domains. The disks discussed in this paper provide, in their dynamics and thermodynamics, a direct bridge between collisionless stellar systems and systems of two dimensional point vortices, the formal analogies between them having been discussed at length in the literature \citep{csr96, chavanis02}.

\subsection{Boundary conditions in the mean field limit}

\noindent We consider boundary conditions on the density $n(\ve)$ near the boundary at $e=1$. To investigate these, we replace $h$ and $k$ in the equations of motion (\ref{eq:hkdot}) by the azimuthal and polar angles on the eccentricity sphere, $\varpi=\hbox{atan2}\,(h,k)$ and $\theta=\cos^{-1}(s\sqrt{1-e^2})$ (eq.\ \ref{eq:thetadef}). Then \begin{equation} {d\varpi\over d\tau}=-2s\frac{\sqrt{1-e^2}}{e^2}\left(k{\p\Gamma\over\p k} +h{\p\Gamma\over\p h}\right),\quad {d\cos\theta\over d\tau}=2\left(h{\p\Gamma\over\p k} -k{\p\Gamma\over\p h}\right).  \end{equation}

In general $\Gamma(\ve)$ is smooth near $e=1$ so the quantities in brackets are also smooth. We conclude that $\dot\theta=\hbox{const}$ and $\dot\varpi\propto s(1-e^2)^{1/2}\to 0$ near $e=1$.  Thus the trajectories intersect the equator ($e=1$) along lines of constant longitude, at constant latitudinal speed.

Since the motion of particles near the equator is smooth, in a steady state the density of particles on the sphere should be smooth near the equator. Since the area element on the eccentricity sphere is proportional to the area element in a canonical phase space, the density of particles on the eccentricity sphere is proportional to the phase-space density $f^+(\vE)$ in the northern hemisphere and $f^-(\vE)$ in the south. Thus $f^+(\vE)$ must join smoothly onto $f^-(\vE)$; in particular $f^+(\vE)=f^-(\vE)$ at the equator. This in turn requires that $n_\pm(\ve)\to c(\varpi)(1-e^2)^{-1/2}$ as $e\to 1$, for some function $c(\varpi)$.

\section{Axisymmetric Keplerian rings and their perturbations}

\subsection{Equilibria}

\label{sec:axi}

\noindent
For axisymmetric disks, the nonlinear Poisson equation governing the mean-field potential of the maximum-entropy disk (eq.\ \ref{eq:thermalde}) simplifies to the ordinary differential equation 
\begin{equation}
\frac{d^2\Psi}{de^2}+\frac{1}{e}\frac{d\Psi}{de}=
\frac{2\alpha}{\sqrt{1-e^2}}\exp[-\Psi(e)]\cosh\gamma\sqrt{1-e^2}.
\label{eq:thermfa}
\end{equation}
To solve this, we write $\Psi(e)=\Psi_0+\psi(e)$, with $\psi(0)=0$, and
define $\overline\alpha\equiv \alpha\exp(-\Psi_0)$. The differential
equation (\ref{eq:thermfa}) becomes 
\begin{equation}
\frac{d^2\psi}{de^2}+\frac{1}{e}\frac{d\psi}{de}=
\frac{2\overline\alpha}{\sqrt{1-e^2}}\exp[-\psi(e)]\cosh\gamma\sqrt{1-e^2};
\label{eq:thermfb}
\end{equation}
for given values of $\overline\alpha$ and $\gamma$, this can be solved
by integrating outwards from $e=0$ with the initial conditions
$\psi(0)=\psi'(0)=0$. The potential at the center is then given by equation
(\ref{eq:thermal}),
\begin{equation}
\Psi_0=2\overline\alpha\int\frac{e\,de}{\sqrt{1-{e}^2}}(\log e -4\log 2)
\exp[-\psi(e)]\cosh\gamma\sqrt{1-{e}^2}.
\label{eq:thermd}
\end{equation}
Knowing
$\psi(e)$ and $\Psi_0$ we can compute the parameter $\alpha$ from
$\overline\alpha$ as well as the dimensionless energy $u$ and angular
momentum $\ell$. 

The differential equation (\ref{eq:thermfb}) does not appear to have a general analytic
solution. However, some aspects of the behavior of these disks 
can be deduced analytically:

\begin{itemize}

\item There is an upper limit to the dimensionless energy $u$: the
  potential $\phi_L(\ve_1,\ve_2)$ increases monotonically with the
  distance $|\ve_1-\ve_2|$ but particles are restricted to
  the circular area $|\ve|\le 1$. Thus the energy of a distribution of
  particles with given mass is maximized if they are uniformly
  distributed on the circle $|\ve|=1$, in which case it is straightforward to show that $u=-2\log 2/\pi=-0.44127$.

\item If the particles have small eccentricities we can replace
  $\sqrt{1-e^2}$ by unity in equation (\ref{eq:thermfb}), to obtain
\begin{equation}
\frac{d^2\psi}{de^2}+\frac{1}{e}\frac{d\psi}{de}=
2\overline\alpha\cosh\gamma\exp[-\psi(e)],
\label{eq:thermff}
\end{equation}
which has the solution \citep{sto63,ost64} 
\begin{equation}
\psi(e)=2\log(1+e^2/e_0^2), \qquad
e_0^2=\frac{4}{\overline\alpha\cosh\gamma}=\frac{4\exp(\Psi_0)}{\alpha\cosh\gamma}. 
\label{eq:ost}
\end{equation}
The validity of this approximate solution requires $e_0\ll1$ and
$\overline\alpha>0$. The dimensionless angular momentum and energy are
\begin{equation}
\ell=\tanh\gamma, \quad u=\frac{1-8\log 2+2\log e_0}{4\pi}.
\label{eq:llldef}
\end{equation}
The parameter $\beta=4\pi$ and the specific entropy is
\begin{equation}
\frac{S}{N}=2-\ell\,\tanh^{-1}\ell+\log(\pi e_0^2) -\log N
-\half\log(1-\ell^2).
\label{eq:sss}
\end{equation}

\item In disks with $\beta=0$ (zero inverse temperature) the dimensionless angular momentum is related to the parameter $\gamma$ by
\begin{equation}
\ell=\coth\gamma-1/\gamma,
\end{equation}
the specific entropy is
\begin{equation}
\frac{S}{N}=1-\gamma\coth\gamma - \log(2\pi\sinh\gamma/\gamma) -\log N;
\end{equation}
and the mean eccentricity and prograde fraction are
\begin{equation}
\langle e\rangle=\frac{\pi I_1(\gamma)}{2\sinh\gamma},\qquad\mbox{prograde}=\frac{\exp(\gamma)-1}{2\sinh\gamma}
\end{equation}
where $I_1$ is a modified Bessel function.

\end{itemize}

Numerical solutions of the differential equation (\ref{eq:thermfb}) are 
discussed in the main text. 

\subsection{Bifurcation to non-axisymmetric disks}

\label{sec:bif}

\noindent
We may use the differential equation (\ref{eq:thermalde}) to
investigate whether the equilibrium axisymmetric disks can remain in equilibrium under small
non-axisymmetric perturbations. If the potential
$\Psi=\Psi_0+\psi_0(e)+\psi_m(e)\exp(im\varpi)$, where $\Psi_0$ and
$\psi_0(e)$ define the potential of the unperturbed axisymmetric
system defined following equation (\ref{eq:thermfa}), $m>0$ is an
integer, and $\psi_m(e)$ is small, equation (\ref{eq:thermalde}) can be
linearized to yield
\begin{equation}
\frac{d^2\psi_m}{de^2}+\frac{1}{e}\frac{d\psi_m}{de}-\frac{m^2}{e^2}\psi_m+
\frac{2\overline\alpha}{\sqrt{1-e^2}}\exp[-\psi_0(e)]\cosh\gamma\sqrt{1-e^2}\,\psi_m=0.
\label{eq:axistab}
\end{equation}
The existence of a solution satisfying the boundary conditions $d\log\psi_m/d\log e=m$ as $e\to 0$ and $=-m$ as $e\to 1$ implies a bifurcation to a sequence of non-axisymmetric disks that initially have $m$-fold symmetry.

Non-axisymmetric equilibria can be stationary in an inertial frame or in a frame precessing with some pattern speed, which we denote $\Omega_p$ (relative to the physical time) or $\omega_p=\Omega_p(2M_\star/M_{\rm disk})(a^3/GM_\star)^{1/2}$ relative to the dimensionless time $\tau$ defined in equation (\ref{eq:ham}).  A rotating, non-axisymmetric equilibrium is a solution of the collisionless Boltzmann equation---which it must be, since the relaxation time is much longer than the orbital or precession time---if and only if the distribution function $f(\vE)$ depends only on the Jacobi integral $J\equiv \widetilde E-\Omega_p\widetilde L$ where $\widetilde E$ and $\widetilde L$ are the non-Keplerian energy and the angular momentum of a single particle. In dimensionless variables $J=(Gm^2N/a)[\Gamma-\half s\omega_p(1-E^2)]$. Comparison to equation (\ref{eq:df}) implies that the dimensionless pattern speed is
\begin{equation}
\omega_p=\frac{2\gamma}{\beta}.
\label{eq:pattern}
\end{equation}

\subsection{Dynamical stability}

\label{sec:dyn}

\noindent
In terms of the Poincar\' e eccentricity $E=(K^2+H^2)^{1/2}$ and the argument of periapsis $\varpi=\tan^{-1}H/K$, the equations of motion (\ref{eq:hameq}) read
\begin{equation}
\frac{dE}{d\tau}=\frac{s}{E}\frac{\p\Gamma}{\p\varpi}, \qquad 
\frac{d\varpi}{d\tau}=-\frac{s}{E}\frac{\p\Gamma}{\p E}.
\end{equation}
The distribution function $f_\pm(\vE,t)$ must satisfy the collisionless Boltzmann equation
\begin{equation}
\frac{\p f_\pm}{\p t} \pm\frac{1}{E}\frac{\p\Gamma}{\p\varpi}\frac{\p f_\pm^0}{\p E}\mp \frac{1}{E}\frac{\p\Gamma}{\p E}\frac{\p f_\pm^0}{\p \varpi}=0.
\label{eq:colbol}
\end{equation}
At this point we switch from the Poincar\'e eccentricity $E$ to the ordinary eccentricity $e=(2E^2-E^4)^{1/2}$:
\begin{equation}
\frac{\p f_\pm}{\p t} \pm2\frac{\sqrt{1-e^2}}{e}\left(\frac{\p\Gamma}{\p\varpi}\frac{\p f_\pm}{\p e}-\frac{\p\Gamma}{\p e}\frac{\p f_\pm}{\p \varpi}\right)=0.
\label{eq:colbol1}
\end{equation}
We now write $\Psi=\beta\Gamma=\Psi_0+\psi_0(e)+\psi_m(e)\exp[i(m\varpi-\omega t)]$, $f_\pm=f^0_\pm(e)+ g^m_\pm(e)\exp[i(m\varpi-\omega t)]$, where $\Psi_0$, $\psi_0(e)$, and $f^0_\pm(e)$ define the potential and distribution function of the unperturbed axisymmetric system, $m>0$ is an integer, and $\psi_m(e)$ and $g_\pm^m(e)$ are small. We then linearize equation (\ref{eq:colbol1}) to obtain
\begin{align}
g^m_\pm(\pm\beta\omega+ 2m\Omega)&=2m\psi_m\frac{\sqrt{1-e^2}}{e}\frac{df_\pm^0}{de} \nonumber \\ &=
-\frac{2mN\overline\alpha}{\beta}\psi_m\exp\big[-\psi_0(e)\pm\gamma\sqrt{1-e^2}\big](\Omega\pm\gamma),
\label{eq:lincolbol}
\end{align}
where $\Omega(e)=\sqrt{1/e^2-1}\,d\psi_0/de$, and the last equality follows from the definition (\ref{eq:df}) of the equilibrium distribution function and $\overline\alpha=\alpha\exp(-\Psi_0)$. 

The perturbed potential $\psi_m$ and the perturbed distribution function $g^m_\pm$ are related by Poisson's equation, which reads
\begin{equation}
\nabla_\ve^2\psi_m=\frac{d^2\psi_m}{de^2}+\frac{1}{e}\frac{d\psi_m}{de}-\frac{m^2}{e^2}\psi_m=\frac{\beta}{N\sqrt{1-e^2}}(g^m_++g^m_-)
\end{equation}
Thus we arrive at an eigenvalue equation for the frequency $\omega$,
\begin{equation}
\frac{d^2\psi_m}{de^2}+\frac{1}{e}\frac{d\psi_m}{de}-\frac{m^2}{e^2}\psi_m+\frac{2\overline\alpha\exp[-\psi_0(e)]}{\sqrt{1-e^2}} \sum_{s=\pm 1} \exp\big(s\gamma\sqrt{1-e^2}\big)\frac{\Omega+s\gamma}{2\Omega+ s\beta\omega/m}\psi_m=0.
\label{eq:dynstab}
\end{equation}

In the special case where the slightly non-axisymmetric system is an equilibrium, $\omega$ is real and equal to $m\omega_p$ where $\omega_p=2\gamma/\beta$ is the pattern speed (eq.\ \ref{eq:pattern}). Then the eigenvalue equation (\ref{eq:dynstab}) reduces to the bifurcation equation (\ref{eq:axistab}). More generally this is a linear differential equation with a nonlinear dependence on the eigenvalue $\omega$. For numerical work it is more convenient to use the integral form of Poisson's equation,
\begin{equation}
\psi_m=-\frac{\beta}{2Nm}\int_0^1 \frac{\epsilon\, d\epsilon}{\sqrt{1-\epsilon^2}}[g^m_+(\epsilon)+g^m_-(\epsilon)]r^m(e,\epsilon)\quad\mbox{where}\quad r(e,\epsilon)=\frac{\mbox{min\,}(e,\epsilon)}{\mbox{max\,}(e,\epsilon)}.
\end{equation}
Together with equation (\ref{eq:lincolbol}) this yields a linear Fredholm integral equation for the perturbed distribution function $g^m_\pm$. After discretization on a grid in $e$, the determination of the eigenvalues $\omega$ reduces to finding the eigenvalues of a real non-symmetric matrix.

\subsection{Entropy of axisymmetric equilibria: maximum or saddle?}

\label{sec:saddle}

\noindent
To establish that an entropy extremum is an entropy maximum we need to
evaluate the changes in entropy due to small changes in the
distribution function from its equilibrium value. Write
$f_\pm(\vE)=f_\pm^0(\vE)+\Delta f_\pm(\vE)$; then to second order in
$\Delta f_\pm$ the changes in number, angular momentum, entropy, and
energy are 
\begin{align}
\Delta N&=\int d\vE\,(\Delta f^++\Delta f^-), \nonumber \\
\Delta L&=m\sqrt{GM_\star a}\int d\vE\,(1-E^2)(\Delta f^+-\Delta f^-),\nonumber \\
\Delta S&=-\Delta N-\int d\vE\,(\Delta f^+\log f_0^++\Delta f^-\log f_0^-)-\int d\vE\left[ \frac{(\Delta f^+)^2}{2f_0^+}+\frac{(\Delta f^-)^2}{2f_0^-}\right]\nonumber \\
\Delta U&=\frac{Gm^2N}{a}\int d\vE \,\Gamma^0(\ve)(\Delta f^++\Delta f^-) \nonumber \\
&\qquad +\frac{Gm^2}{2a}\int d\vE d\vE'\, [\Delta f^+(\vE)+\Delta
f^-(\vE)]\phi(\ve,\ve')[\Delta f^+(\vE')+\Delta f^-(\vE')];\end{align}
here $\Gamma^0(\ve)=\int d\vE'\phi(\ve,\ve')
[f_0^+(\vE')+f_0^-(\vE')]$ and the argument of $f_\pm$ is suppressed
when it is clear from the context. 

We vary the entropy at fixed number, energy, and angular momentum so
$\Delta N=\Delta L=\Delta U=0$. Using these relations and equation
(\ref{eq:df}) we have
\begin{align}
\Delta S&= -\int d\vE\bigg[ \frac{(\Delta f^+)^2}{2f_0^+} +\frac{(\Delta f^-)^2}{2f_0^-}\bigg] \nonumber \\&\qquad -\frac{\beta}{2N}\int d\vE d\vE'\, [\Delta f^+(\vE)+\Delta f^-(\vE)]\phi(\ve,\ve')[\Delta f^+(\vE')+\Delta f^-(\vE')].
\end{align}
This is more conveniently written in terms of functions $\Delta
f\equiv \Delta f^++\Delta f^-$ and $\Delta g\equiv \Delta f^+-\Delta
f^-$:
\begin{align}
\Delta S&=  -\int d\vE\,\frac{[(\Delta f)^2+(\Delta
  g)^2](f_0^++f_0^-)-2\Delta f\Delta g\,(f_0^+-f_0^-)}{8f_0^+f_0^-} \nonumber \\
&\qquad -\frac{\beta}{2N}\int d\vE d\vE'\, \Delta f(\vE)\phi(\ve,\ve')\Delta f(\vE').
\end{align}
The necessary and sufficient condition for stability is that $\Delta
S\le0$ for all variations $\Delta f(\vE)$, $\Delta g(\vE)$. For given $\Delta
f$, it is straightforward to show that $\Delta S$ is maximized when
$\Delta g=\Delta f(f_0^+-f_0^-)/(f_0^++f_0^-)$ so a necessary and
sufficient condition for stability is that 
\begin{equation}
\Delta S=  -\int d\vE\,\frac{(\Delta f)^2}{2(f_0^++f_0^-)}
-\frac{\beta}{2N}\int d\vE d\vE'\, \Delta f(\vE)\phi(\ve,\ve')\Delta
f(\vE').
\label{eq:dsdef}
\end{equation}
is negative or zero for all variations $\Delta f(\vE)$. An equivalent
stability condition is
\begin{equation}
\frac{1}{\lambda}=-\frac{B}{NA}\le 1,
\end{equation}
where
\begin{equation}
A\equiv \int d\vE\,\frac{(\Delta f)^2}{2(f_0^++f_0^-)},\qquad B\equiv 
\frac{\beta}{2}\int d\vE d\vE'\, \Delta f(\vE)\phi(\ve,\ve')\Delta f(\vE').
\end{equation}
The ratio $A/B$ is extremized when $\Delta f$ satisfies 
\begin{equation}
\Delta f+\frac{\lambda\beta}{N}(f_0^++f_0^-)\int d\vE'\Delta
f(\vE')\phi(\ve,\ve')=0.
\label{eq:wpert}
\end{equation}
The equilibrium is stable if and only if all of the eigenvalues $\lambda$ of this
integral equation satisfy $\lambda^{-1}\le 1$.

Equation (\ref{eq:wpert}) can be rewritten as
\begin{equation}
\Delta f+\lambda (f_0^++f_0^-)\Delta\psi=0.
\label{eq:wperta}
\end{equation}
where $\Delta\psi$ is the dimensionless potential due to the density $\Delta f$. 
In common with other sections of this paper, we now replace the potential $\phi(\ve,\ve')$ by the logarithmic potential $\phi_L(\ve,\ve')$ (eq.\ \ref{eq:smalle}). In this case
$\nabla_\ve^2\Delta\psi=\beta N^{-1}\Delta f/\sqrt{1-e^2}$ so the eigenvalue equation simplifies to the differential equation 
\begin{equation}
  \nabla_\ve^2\Delta\psi+ \frac{\lambda\beta}{N}\frac{f_0^++f_0^-}{\sqrt{1-e^2}}\Delta\psi=0.
\label{eq:dea}
\end{equation}

Substituting for the equilibrium distribution function from equation
(\ref{eq:df}), we have
\begin{equation}
\nabla^2_\ve\Delta\psi +\lambda \frac{2\overline\alpha}{\sqrt{1-e^2}}
\exp[-\psi_0(e)]\cosh\gamma\sqrt{1-e^2}\,\Delta\psi=0, 
\label{eq:odeiso}
\end{equation}
where $\psi_0$ is the equilibrium dimensionless potential given by
equation (\ref{eq:thermfb}) and $\overline \alpha$ is defined just
above that equation. 

Writing $\Delta\psi=\psi_m(e)\exp(im\varpi)$ with $m$ a non-negative
integer, we have
\begin{equation}
\frac{d^2\psi_m}{de^2}+\frac{1}{e}\frac{d\psi_m}{de}-\frac{m^2}{e^2}\psi_m+\lambda
\frac{2\overline\alpha}{\sqrt{1-e^2}}\exp[-\psi_0(e)]\cosh\gamma\sqrt{1-e^2}\,\psi_m=0.
\label{eq:thermg}
\end{equation}
The boundary conditions are $d\log\psi_m/d\log e=m$ as $e\to 0$ and
$=-m$ as $e\to 1$. 

This is a Sturm-Liouville equation so the eigenvalues $\lambda$ are real. A necessary condition for stability is that all the eigenvalues satisfy $\lambda^{-1}\le 1$, in other words, $\lambda\le0$ or $\lambda\ge 1$. Moreover the Sturm-Liouville property implies that when $\overline\alpha >0$ there is a minimum eigenvalue $\lambda_0$, so a sufficient condition for stability is $\lambda_0\ge 1$. Similarly, for $\overline\alpha <0$ there is a maximum eigenvalue $\lambda_0$, so a sufficient condition for stability is $\lambda_0\le 0$. Numerical solutions of this equation are discussed in the main text.

In low-eccentricity disks the thermal instability can be described analytically. 
First note that when $e\ll1$ the differential equation (\ref{eq:thermalde}) is $\nabla^2_\ve\Psi=2\alpha\exp(-\Psi)$, which is 
translationally invariant, and hence neutrally stable
to a displacement. At higher order in eccentricity the equation 
contains corrections of order $e^2$ that break the translational
invariance and thus can make modes similar to translations slightly stable or unstable. To
investigate this instability in low-eccentricity disks, we use the
Rayleigh--Ritz variational technique, which states that the minimum
eigenvalue satisfies the inequality
\begin{equation}
  \lambda_0\le \frac{\int_0^1
  de[e{y'}^2(e)+m^2y^2/e]-[ey(e)y'(e)]_{e=0}^1}{2\overline\alpha \int_0^1
  de\,ey^2(e)\exp[-\psi_0(e)]\frac{\displaystyle
    \cosh\gamma\sqrt{1-e^2}}{\displaystyle \sqrt{1-e^2}}}
\end{equation}
for any trial function $y(e)$. The gravitational potential
of a low-eccentricity disk is $\psi_0(e)=2\log(1+e^2/e_0^2)$
(eq.\ \ref{eq:ost}) so the perturbed potential corresponding to a
translation of the disk's center by a small amount $\Delta\ve$ is
$-\psi_0'(e)\ve\cdot\Delta\ve/e=-4e(e_0^2+e^2)^{-1}|\Delta\ve|\cos\varpi$
where $\varpi$ is the angle between $\ve$ and $\Delta\ve$. Thus
a suitable trial function for investigating the stability of $m=1$ modes
similar to translations is $y(e)=e/(e_0^2+e^2)$. In the limit $e_0\ll1$ 
\begin{equation}
\int_0^1 de\,[e{y'}^2(e)+y^2/e]-
[ey(e)y'(e)]_{e=0}^1=\frac{2}{3e_0^2}
+ \mbox{O}(e_0^2) 
\end{equation}
and 
\begin{equation}
2\overline\alpha \int_0^1
  de\,ey^2(e)\exp[-\psi_0(e)]\frac{\cosh\gamma\sqrt{1-e^2}}{\sqrt{1-e^2}}=2\overline\alpha
\cosh\gamma\left[\ffrac{1}{12}+(1-\gamma\tanh\gamma)e_0^2+\mbox{O}(e_0^4)\right].
\end{equation}
Using the relation $e_0^2\,\overline\alpha\cosh\gamma=4$ (eq.\
\ref{eq:ost}) we find
\begin{equation}
\lambda_0\le\left[1+12e_0^2(1-\gamma\tanh\gamma)+\mbox{O}(e_0^4)\right]^{-1}.
\end{equation}
Stability requires $\lambda_0\ge1$ or $\gamma\tanh\gamma\ge 1$, which
in turn requires $\gamma>1.19968$. Since the dimensionless angular
momentum $\ell=\tanh\gamma$ when $e_0\ll1$ (eq.\ \ref{eq:llldef}), an equivalent stability
criterion for low-eccentricity disks is $\ell\ge 0.83356$. As described in the main text, we interpret this finding to imply that axisymmetric equilibrium rings of a given dimensionless angular momentum $\ell$ are stable at all energies when $\ell\ge 0.83356$, while for $\ell < 0.83356$ they are unstable at small energies and low mean eccentricity but stable at high mean eccentricity (the stability boundary is shown in Fig.\ \ref{fig:one}). 

\section{Numerical methods}

\label{sec:num}

\subsection{Optimization over a grid}

\label{sec:method}

\noindent
We work on a grid that is uniformly spaced in the Poincar\'e
eccentricity $E$ and argument of periapsis $\varpi$; between adjacent
grid points $\Delta E=1/M$ and $\Delta\varpi=\pi/M$, typically with 
$M=32$. The distribution function is defined on the space $0<E<1$ and
$0<\varpi<2\pi$; thus there are $2M^2$ grid points and the
distribution function is specified by $4M^2$ variables $f_i^{\pm}$,
which represent the prograde or retrograde phase-space density at grid
point $i$. The potential (eq.\ \ref{eq:pot}) is
defined by its values $\phi_{ij}$ at grid points $i$ and $j$; these
are evaluated by numerical integration once and for all and stored in
a table of size $\sim M^3$.  The phase-space area associated with grid
point $i$ is $A_i=\pi E_i/M^2$. The
total mass, angular momentum, energy, entropy, and 
gravitational potential are evaluated as 
\begin{align}
N&=\sum_i A_i (f^+_i+f^-_i),\nonumber \\
L&=\sum_i A_i(1-E_i^2)(f^+_i-f^-_i),\nonumber \\
U&=\half\sum_i\sum_{j\not=i}
A_iA_j(f_i^++f_i^-)(f_j^++f_j^-)\phi_{ij}+\half\sum_iA_i^2(f^+_i+f^-_i)^2\chi_i=\half\sum_i
A_i(f^+_i+f^-_i)\Gamma_i\nonumber \\
S&=-\sum
A_i[(f_i^++\epsilon)\log(f_i^++\epsilon)+(f_i^-+\epsilon)\log(f_i^-+\epsilon)],\nonumber
\\
\Gamma_i&=\sum_{j\not=i}A_j(f_j^++f_j^-)\phi_{ij}+A_i(f^+_i+f^-_i)\chi_i.
\end{align}
Here $\epsilon$ is a small softening parameter that ensures that the entropy is well-behaved even if the distribution function vanishes at some grid point (typically $\epsilon=0.0001$), and $\chi_i$ is a correction for the self-energy of the material in grid point $i$.

We then find the maximum-entropy state consistent with a given dimensionless energy and angular momentum, using a sequential quadratic programming method (Numerical Algorithms Group, routine E04UCF). The inverse temperature $\beta$ and the pattern speed for non-axisymmetric equilibria are determined by fitting the distribution function to the form $\log f_i^\pm =\mbox{const}-\beta\Gamma_i\pm\gamma(1-E_i^2)$.

\subsection{Optimization using basis functions}

\label{sec:basis}

\noindent
We now describe a basis-function approach to the solution of the integral equation (\ref{eq:thermal}), which gives an independent route to recovering the properties of the entropy extrema using the approximate expression (\ref{eq:potapp}) for the interaction potential. This approach is based on Fourier expansion of the potential followed by an iterative (Nystrom) solution of the integral equation. We first describe the ingredients for axisymmetric solutions, then generalize to non-axisymmetric equilibria.

\paragraph{Axisymmetric solutions}
We will need the cylindrical multipole expansion of the logarithmic kernel:
\begin{align}
\log(\ve-\ve')^2 &= \log[e^2 + e'^2 - 2 e e' \cos(\Delta\varpi)] \nonumber \\
                           &= \log(e_>^2) - 2\sum_{k=1}^\infty \frac{1}{k} \left(\frac{e_<}{e_>}\right)^k \cos(k\Delta\varpi)
\label{eq:kerngt}
\end{align}
where $\Delta\varpi=\varpi-\varpi'$, $e_<=\mbox{\,min}(e,e')$ and $e_>=\mbox{\,max}(e,e')$. 

We write the interaction potential in the form $\phi(\ve, \ve')= {\phi}_{a} + {\phi}_{b}\log(\ve-\ve')^2$ where
\begin{align}
\phi_a&=a_0+ a_1 (e^2 + e'^2) + a_2 e e' \cos \Delta \varpi + a_3 e^2 e'^2 +a_4 (e^4 + e'^4) \nonumber \\ 
           &+ a_5(e^2+e'^2) e e' \cos \Delta \varpi + a_6 e^2 e'^2 \cos^{2} \Delta \varpi  \nonumber  \\
 \phi_b&=b_0 + b_1 (e^2 + e'^2) + b_2 e e'\cos\Delta\varpi.
\end{align}
and the coefficients $a_i$, $b_i$ are given in equation (\ref{eq:fit}). In axisymmetric solutions $\Psi(e)$ is independent of $\varpi$, so the integration over $\varpi'$ in equation  (\ref{eq:thermal}) can be done explicitly to yield:
\begin{equation}
\Psi(e) =4 \pi \alpha{\int}_{0}^{1} {\phi}_{\rm axi} (e, e') \exp[-\Psi(e')] \cosh\gamma(1-{E'}^2) E' dE',  
\label{eq:inteq}
\end{equation}
with
\begin{align}
{\phi}_{\rm axi} (e, e') & = a_0+ a_1 (e^2 + e'^2) + (a_3+\half a_6) e^2 e'^2 +a_4 (e^4 + e'^4)  \nonumber \\ &+ \left\{
\begin{array}{c} 2 [b_0 + b_1 (e^2 + e'^2)]\log(e) - b_2 e'^2,  \quad e > e' \\
2 [b_0 + b_1 (e^2 + e'^2)]\log(e') -  b_2 e^2, \quad e<e'.\end{array} \right \}
\end{align}

To recover the axisymmetric solutions of the integral equation (\ref{eq:thermal}) we discretize the integral over $E$ using Gauss-Legendre abscissae, perform the quadrature with the appropriate weights over the interval $[0,1]$,  and iterate for given values of $\alpha$ and $\gamma$ until we converge to a solution. 

One issue to be dealt with is numerical instabilities, which appear to arise in the constant term in ${\phi}_a$. These can be suppressed by eliminating the variables $\Psi(e)$ and $\alpha$ in favor of $\psi(e)=\Psi(e)-\Psi(0)$ and ${\overline \alpha} = \alpha \exp(-{\Psi}_{0})$. We have 
\begin{equation}
{\psi}(e)=4 \pi {\overline \alpha} {\int} [{\phi}_{\rm axi}(e, e') -{\phi}_{\rm axi}(0, e') ]\exp[-{\psi}(e')] \cosh\gamma(1-{E'}^2) E' dE.
\end{equation}
One then solves for $\psi(e)$ having specified ${\overline \alpha}, \gamma$; then one recovers
\begin{equation}
 {\Psi}_{0} = 4 \pi {\overline \alpha} {\int} {\phi}_{\rm axi}(0, e')  \exp[-{\psi}(e')] \cosh\gamma(1-{E'}^2) E' dE'
\end{equation}
and $\alpha ={\overline \alpha} \exp({\Psi}_{0})$. 

\paragraph{Non-axisymmetric solutions}
It is straightforward to generalize the treatment above to non-axisymmetric solutions. Instead of a single integral equation, there is now a set of coupled integral equations. Recall that
\begin{equation}
\exp[a \cos(\theta)] = I_0 (a) + 2 \sum_{k=1}^{\infty} I_{k}(a) \cos(k \theta),
\end{equation}
where $I_{k}$ is a modified Bessel function. Writing
\begin{equation}
\Psi(e, \varpi) = {\Psi}_0 (e) +  \sum_{k=1}^{\infty} {\Psi}_{k}(e) \cos(k \varpi),
\end{equation}
we work our way through the integral equation to recover a system of integral equations coupling the mode amplitudes $ {\Psi}_{k}$ up to the desired order. Here, we illustrate the process using only $k=0$ and $k=1$, with the understanding that the treatment can be generalized to arbitrary order. 
We have 
\begin{eqnarray}
\exp(-\Psi) &= & \exp (-{\Psi}_{0}) \exp-({\Psi}_{1} \cos\varpi) \nonumber \\
                  &= & \exp (-{\Psi}_{0})\bigg[ I_0 (-{\Psi}_{1}) + 2 \sum_{k=1}^{\infty} I_{k}(-{\Psi}_{1}) \cos(k \varpi) \bigg];
\end{eqnarray}
truncating the sum at $k=1$ and using the relation $I_k(-x)=(-1)^kI_k(x)$ gives 
\begin{equation}
\exp(-\Psi) \simeq  \exp(-{\Psi}_{0})\left [ I_0 ({\Psi}_{1}) - 2\,  I_{1}({\Psi}_{1}) \cos( \varpi) \right ].
\end{equation}
Inserting this result into the integral equation (\ref{eq:inteq}) and integrating over ${\varpi}'$, we end up with two integral equations for ${\Psi}_{0}$ and ${\Psi}_{1}$: 
\begin{eqnarray}
{\Psi}_{0} & = & 4 \pi \alpha \int {\phi}_{0} (e, e') \exp [-{\Psi}_{0}] \,I_{0}({\Psi}_{1}) \cosh\gamma(1-{E'}^2)\,E' dE' \\ \nonumber
{\Psi}_{1} & = & 4 \pi \alpha \int {\phi}_{1} (e, e') \exp [-{\Psi}_{0}]\, I_{1}({\Psi}_{1})  \cosh\gamma(1-{E'}^2)\,E' dE',
\end{eqnarray}
where $\phi_{0}$ and ${\phi}_{1}$ have contributions from both ${\phi}_{a}$ and ${\phi}_{b}$:
\begin{equation}
\phi_0^{a} = a_0+ a_1 (e^2 + e'^2) + (a_3 + \half a_6)e^2e'^2 +  a_4 (e^4 + e'^4) ,\qquad \phi_1^{a} = -a_2 e e' - a_5 (e^2+e'^2) e e',
\end{equation}
and 
\begin{equation}
\phi_0^{b} = \begin{array}{c} 2 [b_0 + b_1 (e^2 + e'^2)]\log(e) - b_2 e'^2,  \quad e > e' \\
2 [b_0 + b_1 (e^2 + e'^2)]\log(e') -  b_2 e^2, \quad e<e'\end{array},
\end{equation}
\begin{equation}
\phi_1^{b} = \begin{array}{c} -2 b_2 e e' \log(e) + \half b_2{e'}^3/e + 2 [b_0 + b_1 (e^2 + e'^2)] (e'/e),  \quad e > e' \\
- 2 b_2 e e' \log(e') + \half b_2e^3/e'+ 2 [b_0 + b_1 (e^2 + e'^2)] (e/e'), \quad e<e'. \end{array}
\end{equation}
Once again, for numerical stability it is better to work with $\psi_k(e)$ and $\overline\alpha$ rather than $\Psi_k(e)$ and $\alpha$. 

Equilibria with pure logarithmic interactions can be recovered by taking $a_0 = -4 \log 2/\pi$, $b_0 = 1/2\pi$ and setting the remaining $a_i$ and $b_i$ to zero.

\bibliography{touma-tremaine-rev-ms}

\end{document}